\documentclass[twocolumn]{aastex62}
\usepackage{float}
\usepackage{gensymb}
\usepackage{amsmath}

\usepackage{comment}
\usepackage{subfigure}
\usepackage{enumitem}
\graphicspath{{./}{figures/}}

\shorttitle{Pulsar Caps}
\shortauthors{Lockhart et al.}

\begin{document}

\title{X-ray Lightcurves from Realistic Polar Cap Models: Inclined Pulsar Magnetospheres and Multipole Fields}

\author{Will Lockhart}
\affiliation{Department of Physics, University of Arizona, Tucson, AZ 85721, USA}

\author{Samuel E. Gralla}
\affiliation{Department of Physics, University of Arizona, Tucson, AZ 85721, USA}

\author{Feryal {\"O}zel}
\affiliation{Department of Astronomy and Steward Observatory, University of Arizona, 933 N. Cherry Avenue, Tucson, AZ 85721, USA}

\author{Dimitrios Psaltis}
\affiliation{Department of Astronomy and Steward Observatory, University of Arizona, 933 N. Cherry Avenue, Tucson, AZ 85721, USA}

\begin{abstract}
Thermal X-ray emission from rotation-powered pulsars is believed to originate from localized ``hotspots'' on the stellar surface occurring where large-scale currents from the magnetosphere return to heat the atmosphere.  Lightcurve modeling has primarily been limited to simple models, such as circular antipodal emitting regions with constant temperature.  We calculate more realistic temperature distributions within the polar caps, taking advantage of recent advances in magnetospheric theory, and we consider their effect on the predicted lightcurves.  The emitting regions are non-circular even for a pure dipole magnetic field, and the inclusion of an aligned magnetic quadrupole moment introduces a north-south asymmetry.  As the quadrupole moment is increased, one hotspot grows in size before becoming a thin ring surrounding the star.  For the pure dipole case, moving to the more realistic model changes the lightcurves by $5-10\%$ for millisecond pulsars, helping to quantify the systematic uncertainty present in current dipolar models.  Including the quadrupole gives considerable freedom in generating more complex lightcurves. We explore whether these simple dipole+quadrupole models can account for the qualitative features of the lightcurve of PSR J0437$-$4715.
\end{abstract}


\section{Introduction} \label{sec:intro}

Since their discovery in the 1960s, pulsars have offered a unique window onto the exotic physics of neutron stars~\citep[for a recent review, see][]{Ozel2016}.  Radio pulsars that are gravitationally bound to a companion in a binary orbit lead to precise measurements of the neutron-star masses and to global constraints on the properties of ultra-dense baryonic matter \citep{Demorest2010,Antoniadis2013,Antoniadis2016,Fonseca2016}. Observations of pulsar glitches and of the rate of thermal cooling of young pulsars offer additional information about their moments of inertia \citep{Link1999} and their internal compositions \citep{Page2006}, respectively. Spectroscopic analyses of the thermal emission from bursting and quiescent neutron stars have provided direct measurements of the radii of a handful of sources \citep{Ozel2009a,Guver2010,Ozel2010,Steiner2010,Guillot2013,Guillot2014,Heinke2014,Nattila2016,Ozel2016a,Bogdanov2016}. More recently, the detection of gravitational waves from the neutron-star merger event GW170817 placed constraints on the tidal deformability \citep{Abbott2017,Abbott2018} and, consequently, on the radii of the two neutron stars~\citep{Raithel2018,De2018,Zhao2018,Abbott2018,Radice2018,Coughlin2018}. 

A complementary approach to measuring neutron-star radii involves modeling the lightcurves of the surface thermal emission from rotation-powered millisecond pulsars.  This emission is thought to arise from localized ``hotspots'' on the stellar surface that come in and out of view as the star rotates, causing the observed brightness to oscillate periodically in time.  The self-lensing of this emission in the gravitational field of a neutron star of mass $M$ and radius $R$ suppresses the amplitude of pulsations by factors that depend primarily on the compactness of the neutron star $GM/Rc^2$, where $G$ and $c$ are the gravitational constant and the speed of light, respectively \citep{Pechenick1983}. As a result, fitting the detailed properties of X-ray lightcurves observed from millisecond pulsars of known mass can lead, in principle, to a measurement of their radii~\citep[see, e.g.,][]{Zavlin1995,Zavlin1998,Bogdanov2007,Bogdanov2013,Psaltis2014a,Ozel2016,Watts2016,Bogdanov2019}. 

The Neutron-star Interior Composition ExploreR (NICER) is now measuring X-ray lightcurves from millisecond pulsars with sensitivity that is poised to make such measurements possible \citep{Gendreau2016}. However, to exploit the full capabilities of these observations, a new generation of theoretical modeling is required to extract physical parameters from these lightcurves and thereby advance efforts to constrain the neutron-star equation of state.

In the canonical pulsar model, the neutron star surface is heated by relativistic electrons and positrons arising from the pair production that supports the global magnetospheric circuit \citep{Arons1981,Arons1983}.  The current flows along ``open" field lines (i.e., those field lines that cross the light cylinder), whose intersection with the stellar surface defines the polar caps.\footnote{Here, the light cylinder $R_{\rm L}\equiv c / \Omega$ is defined as the distance at which an object co-rotating with a star with angular spin frequency $\Omega$ would move at the speed of light.  Note that we will retain the term ``polar caps'' even when these regions are neither polar nor cap-shaped.} For a pure dipole field configuration, the polar caps are circles centered on the magnetic poles (at least at slow rotation), but more generally a variety of shapes are allowed.  Furthermore, the bombardment (and thus the resulting emission) occurs over only a sub-region of the polar caps determined by the detailed properties of the current \citep{Timokhin2013}.  In this framework, the configuration and multipolar structure of the magnetic field of the neutron star determines the properties of the emitting regions.

Past efforts to model the thermal surface emission from pulsars have usually used circular hotspots that coincide with the magnetic poles under the implicit assumption of a purely dipolar magnetic field~\citep[see, e.g.,][]{Zavlin1998}.  This canonical choice was driven by the relative (numerical) simplicity of simulating the dipolar configuration as well as the lack of an obvious, physically motivated alternative.  However, there is little reason to believe that pulsar fields are purely dipolar. Indeed it is natural to suspect that the complexity of pulsar emission, and its variation across sources~\citep{Lorimer2008,Rankin2017}, are driven by similar complexity and variation in the pulsar magnetic field. Furthermore, for X-ray pulse profile modeling itself, the lightcurves and spectra observed from the prototypical source PSR~J0437$-$4715 have required multiple concentric regions of emission with different temperatures, as well as two hotspots of emission that are not antipodal~\citep{Bogdanov2013}.  On the theoretical side, recent force-free and particle-in-cell (PIC) simulations~\citep{Spitkovsky2006, Bai2010,Chen2014,Philippov2015a,Philippov2015b,Brambilla2018} and analytic calculations~\citep{Gralla2016,Belyaev2016,Gralla2017} have revealed a similar degree of complexity in polar cap properties.

In this paper, we model the shapes and thermal properties of polar caps on neutron stars with dipolar and ``quadrudipolar" (i.e., dipole+quadrupole) magnetic fields using earlier semi-analytic general-relativistic calculations of force-free currents in their magnetospheres~\citep{Gralla2017}. We combine these calculations with basic models of the resulting surface emission and with ray-tracing calculations in the external spacetimes of the neutron stars~\citep[see, e.g.,][]{Baubock2012,Baubock2013,Psaltis2014b} to predict the expected lightcurves from these polar-cap models. Our goal is to explore new pulse morphologies and properties that result from more realistic polar-cap models and understand the magnitude of potential biases that may be introduced by the simplified configurations commonly used.

We find that realistic polar-cap models naturally give rise to regions of surface emission that have different temperatures and to lightcurves with primary and secondary peaks that are not offset by 180 degrees, as also inferred from observations. Furthermore, the realistic polar-cap models introduce complexity in the resulting lightcurves at the $5-10$\% level. These effects will mask and alter similar complexity in the lightcurves introduced by rotational Doppler effects that depend primarily on the neutron-star radii~\citep{Psaltis2014a}. As a result, not accounting for the thermal structure of the pulsar polar caps can lead to imprecise model selection and may bias the radii measurements obtained with this approach.

\section{Method} \label{sec:method}

In this section, we introduce the various ingredients of our method, beginning with a set of neutron star parameters and arriving at a simulated lightcurve. Throughout the paper, we consider a neutron star of mass $M$ and radius $R$, rotating with angular frequency $\Omega$.  We denote polar coordinates around the spin axis of the star by the colatitude $\theta$ and azimuth $\phi$, and the inclination of the observer's line of sight with respect to the spin axis by $\theta_o$. We consider the case where the magnetic field of the neutron star is axisymmetric with respect to a ``magnetic axis'' that co-rotates with the star at an inclination angle $\zeta$ relative to the axis of rotation. We also set $G=c=1$, unless explicitly stated otherwise. 

Our procedure is composed of three steps:
\begin{enumerate}
    \item From a choice of stellar parameters and magnetic field
      configuration, we compute the four-current density $J^\mu(\theta,\phi)$ near
      the star using the model of a slowly rotating, perfectly
      conducting star with a force-free magnetosphere~\citep{Gralla2017}.
    \item Given the current $J^\mu(\theta,\phi)$ near the star, we
      determine the surface temperature $T(\theta,\phi)$ via different
      phenomenological prescriptions for Joule heating.  From this
      temperature, we then determine the specific intensity of the emerging radiation using a blackbody spectrum and a simple beaming function which we define later on. 
    \item Finally, we simulate the predicted lightcurve using 
      general-relativistic ray-tracing of null geodesics from the stellar surface to an observer at infinity~\citep{Baubock2012,Baubock2013,Psaltis2014b}.
\end{enumerate}
Steps 1 and 3 rest on well-established underlying physics and
contribute little systematic uncertainty to the results.  By contrast, the phenomenological prescriptions in step 2 are motivated but by no means definitive. We therefore pay careful attention to the dependence of the lightcurves on the choice of prescription.  To compare with observations would require a further step 4 of taking into account foregrounds and instrumental response, which is beyond the scope of the current work.

\subsection{From Magnetization to Current}

Our first task, given the parameters of the star and its magnetic
field configuration, is to determine the current flow at the surface.
This flow is sensitive to the global structure of the magnetospheric
circuit (extending past the light cylinder), which can in general be
determined only with expensive numerical simulations. However, in the slow-rotation limit, there is a decoupling of near and far regions that enables the full general-relativistic, non-dipolar problem to be solved analytically with input from a \textit{single} simulation of a dipole pulsar in flat spacetime \citep{Gralla2017}.  

We follow Gralla et al. and keep terms to leading order in the rotation parameter $\Omega R / c$.  For NICER's main target, and targets of similar interest, this parameter is estimated to be $\Omega R / c \sim 0.05$. By keeping to leading order in rotation one can therefore expect relative corrections of $(0.05)^2$ from higher-order terms, which can safely be ignored. We briefly review this approach and present the formulae we will need for lightcurve modeling.

To leading order in rotation, the spacetime geometry outside of a general-relativistic star is given by the linearized Kerr metric,
\begin{align}\label{eq:StarMetric}
\begin{split}
    ds^2&=-f(r)dt^2+f(r)^{-1} dr^2 \, + \\ 
    &\quad r^2\left[d\theta^2+\sin^2{\theta}\left(d\phi-\Omega_Z(r) dt\right)^2\right],\quad	r>R,
\end{split}
\end{align}
where $f(r)=1-2M/r$ is the square of the redshift factor, $\Omega_Z=2 I \Omega/r^3$ is the frame-dragging frequency, and $I$ is the moment of inertia (angular momentum over angular velocity).  Note that the surface of the star is spherically symmetric at this order, although the spacetime is only axisymmetric.

Assuming a perfectly conducting star, the force-free magnetosphere is completely determined by specifying the radial component of the magnetic field on the stellar surface, $B^r_\star(\theta,\phi,t)$. Let \textit{primed} angles $(\theta',\phi')$ denote polar coordinates with respect to the magnetic axis, related to the unprimed coordinates by a rotation,\footnote{We work at an instant of time and make the arbitrary choice to align the angular velocity with the $z$-axis.}
\begin{subequations}
    \begin{align}
        \cos{\theta'} &= \cos{\theta}\cos{\zeta} - \sin{\theta}\cos{\phi}\sin{\zeta}, \\
        \tan{\phi'} &= \frac{\sin{\theta}\sin{\phi}}{\sin{\theta}\cos{\phi}\cos{\zeta} + \cos{\theta}\sin{\zeta}}.
    \end{align}
\end{subequations}
Since the field is axisymmetric, we can express $B^r_\star$ in terms of the magnetic flux in the co-rotating frame, $\psi_\star(\theta')=\int B^r_\star \sqrt{-g} d\theta'$. We may further decompose this flux into a set of multipole moments $B_\ell$,

\begin{equation}
\label{eq:moments}
    \psi_\star(\theta') = R^2 \sum_{\ell=1}^\infty B_{\ell} \, \Theta_\ell(\theta'),
\end{equation}
where $\Theta_\ell$ are the eigenfunctions for magnetic flux~\citep{Gralla2016}. To leading order in $\Omega R$, the magnetosphere is completely determined by the choice of parameters $\{ M, R, I, \Omega, \zeta, B_\ell \}$ defined above. 

Using this framework, \citet{Gralla2017} have shown that the four-current density $J^\mu$ near the star has the following orthonormal frame components:
\begin{subequations}\label{eqn:4}
    \begin{align}
        \begin{split}
            J^{\hat{t}} &= \frac{\Omega - \Omega_Z}{\sqrt{f}r^2} \bigg[ \partial_\theta \alpha \, \partial_\theta (\partial_\phi \beta) - \partial_\theta \beta \, \partial_\theta (\partial_\phi \alpha) - \partial_\phi \alpha \\
            & \left( \frac{\partial_\theta (\sin{\theta} \partial_\theta \beta)}{\sin{\theta}} \right) + \partial_\phi \beta \left( f \partial_r (r^2 \partial_r \alpha) + \frac{\partial_\theta (\sin{\theta} \partial_\theta \alpha)}{\sin{\theta}} \right)\! \bigg]
        \end{split} \\
        J^{\hat{r}} &= \frac{\Lambda(\alpha,\beta)}{\sqrt{f} \, r^2 \sin{\theta}} (\partial_\theta \alpha \, \partial_\phi \beta - \partial_\theta \beta \, \partial_\phi \alpha )  \\
        J^{\hat{\theta}} &= - \frac{\Lambda(\alpha,\beta)}{r \sin{\theta}} (\partial_r \alpha \, \partial_\phi \beta) \\
        J^{\hat{\phi}} &= \frac{\Lambda(\alpha,\beta)}{r} (\partial_r \alpha \, \partial_\theta \beta),
    \end{align}
\end{subequations}
where all the terms are to be evaluated at $r=R$. The scalars $\alpha$ and $\beta$ are ``Euler potentials'' for the magnetic field near the star, given explicitly by   
\begin{align}\label{alphabeta}
    \alpha(r,\theta,\phi) &= R^2 \sum_{l=1}^\infty B_l \, R_\ell^>(r) \Theta_l(\theta'), \\ \beta(r,\theta,\phi) &= \phi',
\end{align}
where $R_\ell^>(r)$ are radial eigenfunctions (\citet{Gralla2016} and Eq.~\ref{eq:R2} below). Finally, $\Lambda(\alpha,\beta)$ is a conserved quantity on field lines related to the current on each line.  \citet{Gralla2017} provide an analytic expression for $\Lambda(\alpha,\beta)$, determined by fitting numerical data, in terms of Bessel functions $J_n$,\footnote{$\mp$ refers to the north/south hemisphere. The internal $\mp$ appearing in front of $J_1$ was missing from the original formula in \citet{Gralla2017}, and has been corrected here.} 
\begin{align}
    \Lambda(\alpha,\beta) = \begin{cases} \, \mp 2 \Omega \bigg[ J_0 \left( 2 \arcsin{\sqrt{\alpha/\alpha_0}} \right) \cos{\zeta} \\ 
    \quad \mp J_1 \left( 2 \arcsin{\sqrt{\alpha/\alpha_0}} \right) \cos{\beta}\sin{\zeta} \bigg], & \alpha < \alpha_0 \\
    0, & \alpha > \alpha_0
    \end{cases},
\end{align}
with the boundary of the polar cap delineated by
\begin{align}
    \alpha_0 &= \sqrt{3/2} \, \mu \Omega \left( 1 + \frac{1}{5} \label{alpha0} \sin^2\zeta \right), \\
    \mu &= B_1 \frac{R^2}{R^{>}_1(R)}.\label{compact}
\end{align}
Here $\mu$ is the magnitude of the dipole moment at the light cylinder, which Eq.~\eqref{compact} relates to the dipole moment $B_1$ at the star by compactness-dependent factors given below (see Eq.~\ref{R1}). The polar caps are defined by $0<\alpha<\alpha_0$.  Their physical size and shape follows from the magnetic field, the magnetic inclination, the rotation rate, the moment of inertia, and the compactness via Eq.~\eqref{alphabeta} and Eq.~\eqref{alpha0}.  The area of the polar cap scales as $(\Omega R)^2$ when the other parameters are held fixed.  In this paper, we focus on the effects of magnetic field and inclination, setting the compactness and moment of inertia to fiducial values given in \S~\ref{emerging}.

These expressions allow the current $J^\mu$ on the star to be computed analytically from the parameters $\{ M, R, I, \Omega, \zeta, B_\ell \}$.  
\begin{figure*}[t]
\centering
\includegraphics[width=\textwidth]{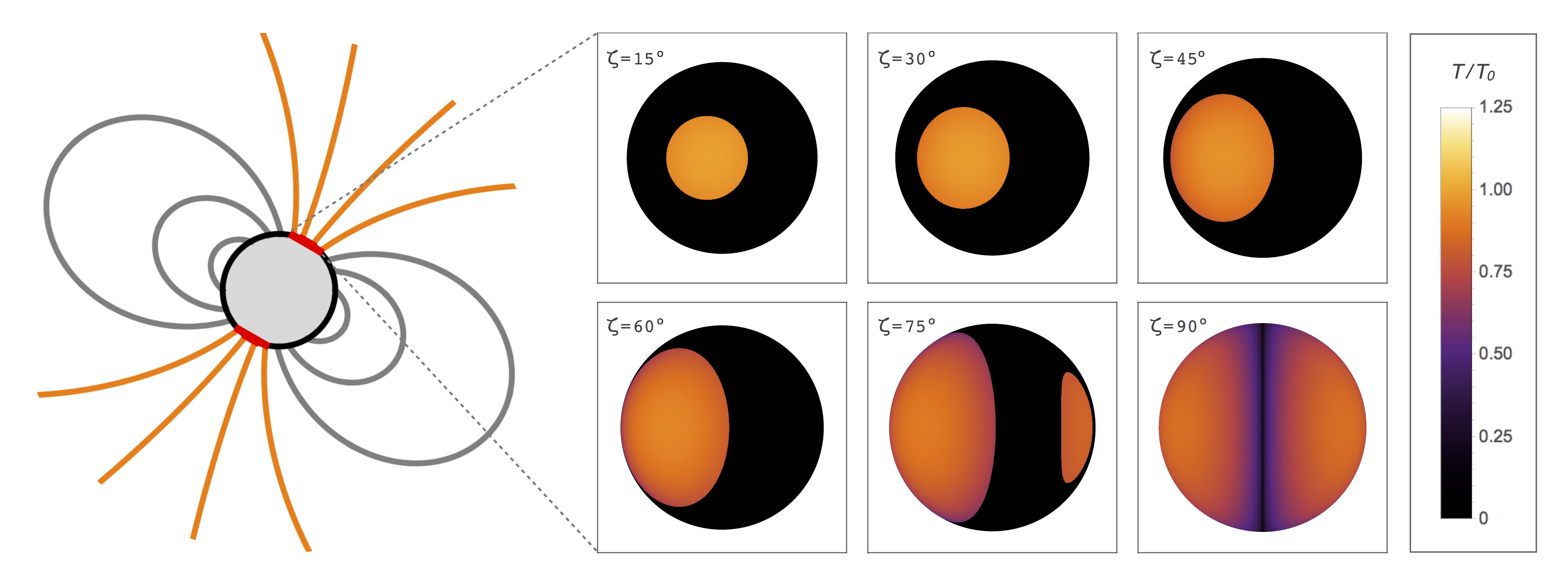}
\caption{{\em (Left)\/} Closed (gray) and open (orange) field lines along which current flows onto a neutron star with a dipolar magnetic field (the spin axis is oriented vertically). {\em (Right)\/} The temperature distribution on the northern polar cap, shown for different inclinations of the magnetic dipole ranging from $\zeta = 15\degree$ (top left) to $\zeta = 90\degree$ (bottom right). The area of the polar caps scales with $(\Omega R)^2$. For our canonical parameters, the half-opening angle of the polar cap lies in the range $14 - 16$ degrees, depending on the magnetic inclination. A second, disjoint emitting lobe forms within the polar cap when the magnetic inclination exceeds $\zeta_{\rm{crit}} \simeq 70\degree$.  \label{fig:fig1}}
\vspace{10pt}
\end{figure*}

\subsection{From Current to Emerging Radiation}

The mechanism most commonly invoked to explain the thermal surface emission from millisecond pulsars is the bombardment of the stellar atmosphere by relativistic electrons and positrons, produced in so-called ``gaps'' (regions of non-zero $\mathbf{E}\cdot\mathbf{B}$) just outside the star and accelerated by the fields \citep{Arons1981,Arons1983}.  While current flows over the entirety of the polar cap (by definition), pair production (leading to bombardment) occurs mainly in regions where the four-current in the corresponding force-free solution is spacelike ($J_\mu J^\mu>0$), which requires the presence of both signs of charge \citep{Timokhin2013,Philippov2015a}. This spacelike region defines the shape of the hotspot in our model.\footnote{Pair production is also expected to occur in the thin current sheet on the last open field lines, and possibly in nearby regions of volume return current  \citep{Timokhin2013, Philippov2018}. This could give rise to further X-ray emission from near the edges of the polar cap, an effect we neglect at present.} 

The detailed plasma physics governing the deposition of energy by magnetospheric currents in the surface layers of the neutron star is explored in~\citet{Baubock2019}. As electrons and positrons bombard the surface of the star, their kinetic energy is deposited onto a thin layer of the stellar atmosphere. Assuming that a fixed fraction $\chi$ of the total current $|\vec{J}|$ is carried by relativistic particles with typical Lorentz factor $\bar{\gamma}$ traveling towards the stellar surface, we write the rate at which this energy is deposited in the surface layers as
\begin{equation}
    P_{\rm{in}} =  (\bar{\gamma} - 1)m_{\rm e} \chi \frac{\vert\vec{J}\vert}{e}\;,
\end{equation}
where $e$ and $m_{\rm e}$ are the electron charge and mass, respectively. The rate at which this energy is deposited must equal the power radiated, which can be written in terms of the effective temperature using the Stefan-Boltzmann law as $P_{\rm{out}} = \sigma T^4$. Setting $P_{\rm{in}} = P_{\rm{out}}$ gives 
\begin{equation}
   T= \left[\frac{m_e \chi (\bar{\gamma}-1)}{e \sigma} \right]^{1/4} |\vec{J}|^{1/4} \;.
\end{equation}
Using this scaling and neglecting, for simplicity, the details of the atmospheric structure leads to the simple prescription
for the effective temperature on the polar cap
\begin{equation}
\label{eq:TJ4}
	T(\theta,\phi) \equiv \begin{cases} 
     \, T_0|\vec{J}/J_0|^{1/4}, & |\rho| < |\vec{J}| \quad (\textrm{spacelike}) \\
     \, 0, & |\rho| > |\vec{J}| \quad (\textrm{timelike})
  \end{cases},
\end{equation}
where $|\vec{J}|= \sqrt{(J^{\hat{r}})^2+(J^{\hat{\theta}})^2+(J^{\hat{\phi}})^2}$ is the three-current density, $\rho=J^{\hat{t}}$ is the charge density, $J_0$ is the maximum value of $|\vec{J}|$ for a neutron star with an aligned dipolar magnetic field, and $T_0$ is the corresponding maximum effective temperature for the aligned dipole.

In order to convert the effective temperature to a model for the emerging specific intensity of the radiation field, we need, in principle, a detailed calculation of the thermal structure of the neutron-star atmosphere~\citep[as done, e.g., in][]{Baubock2019}. This is beyond the scope of this paper. Instead, we will use a simplified approximation of blackbody emission at the effective temperature, with beaming that is potentially non-isotropic,
 \begin{equation}\label{blackbody}
	I(\theta,\phi;\epsilon,\Theta) = \frac{\epsilon^3}{\exp[\epsilon/T(\theta,\phi)]-1} \mathcal{B}(\Theta).
\end{equation}
In this last equation, we denote the photon energy by $\epsilon$, the angle to surface normal by $\Theta$, and the beaming function by $\mathcal{B}(\Theta)$. We will mainly consider the case of isotropic emission ($\mathcal{B}=1$) except in \S\ref{sec:J0437}, where we introduce a simple model for limb-darkening, $\mathcal{B}=1 + h \cos^2\Theta$, with $h$ a free parameter.
Equation~\eqref{blackbody} represents a compromise between realism and simplicity, given current understanding of the underlying micro-physical processes.  Ultimately we expect that a more accurate prescription will be determined by simulations; in the meantime, we estimate the importance of the precise prescription by comparing to other reasonable models (\S\ref{sec:dipole} below).

\subsection{From Emerging Radiation to Lightcurves} \label{emerging}

Using equation~\eqref{blackbody} as a boundary condition, we then use the numerical algorithm of~\citet{Psaltis2014b} to trace the light rays along the null geodesics of the spacetime and map the surface of the simulated compact object onto the image plane of an observer at infinity.  We keep only terms that are up to first order in the neutron-star spin frequency, treat the star as spherical \citep[i.e., neglecting any mass quadrupole terms; see, e.g.,][]{Baubock2013}, and set the moment of inertia to $(2/5)MR^2$ for simplicity.

The simulations presented in this paper have eight parameters: the mass $M$, radius $R$, and spin frequency $f_{\rm s}$ of the star (or equivalently, the angular frequency $\Omega=2\pi f_{\rm s}$); the inclination angle $\theta_o$ between the spin axis and the line of sight of the observer; the inclination angle $\zeta$ of the magnetic axis from the spin axis; the beaming factor $h$; the temperature constant $T_0$, which relates current density to temperature; and the magnetic quadrupole-to-dipole ratio $q$. When comparing to observations of a known millisecond pulsar, the spin frequency is known a priori.

In the following sections we present analysis of lightcurves from pulsars with non-negligible dipolar and quadrupolar components of the magnetic fields. The procedure to include higher-order terms is straightforward in principle. We take as our default choice of stellar parameters $M=1.5M_\odot, R=11$~km, and $f_{\rm s} = 300~$Hz, which corresponds to a dimensionless compactness $M/R \simeq 0.2$ and surface velocity $\Omega R \simeq 0.07c$.  Unless stated otherwise, all plots are shown for these fiducial values.

\section{Pulsars with Pure Dipolar \\ Magnetic Fields} \label{sec:dipole}

The simplest non-trivial magnetic field configuration is a pure dipole, where only the dipole moment $B_1$ in Eq.~(\ref{eq:moments}) is non-zero.  The eigenfunctions for magnetic flux reduce to\footnote{Note that the expression for $R^{>}_1$ present in \citet{Gralla2017} contains a typographical error.}
\begin{align} 
       \Theta_1(\theta) &= \sin^2 \theta, \\
       R^{>}_1(r) &= \left(\frac{3}{2r}\right)\frac{3-4f+f^2+2\log{f}}{(1-f)^3},
       \label{R1}
\end{align}
where we remind the reader that $f=(1-2M/r)$. Equation~\eqref{alphabeta} for the Euler potentials becomes
\begin{equation}\label{alphabeta-dipole}
        \alpha = B_1 R^2 \left( \frac{R^{>}_1(r)}{R^{>}_1(R)} \right) \sin^2{\theta'}, \quad
        \beta = \phi'.
\end{equation}
Substituting equation~\eqref{alphabeta-dipole} into equation~\eqref{eqn:4} gives the current flow on the star, from which equation~\eqref{eq:TJ4} provides the temperature distribution.  The polar caps are antipodally symmetric.  

Figure \ref{fig:fig1} shows the shape and temperature distribution on the polar caps for six different magnetic inclination angles.  For our canonical parameters, the half-opening angle of the polar cap lies in the range $14 - 16\degree$ depending on the inclination. For small inclinations, the emitting region (i.e., the region with spacelike currents) is smaller than the polar cap but centered on the magnetic axis. As the inclination increases, the emitting region becomes larger and its center is displaced from the magnetic axis (i.e., the center of the cap) until, at a critical angle ($\zeta \simeq 70\degree$ for the fiducial parameters), a second lobe forms. 

\begin{figure}
\centering
\includegraphics[width=\linewidth]{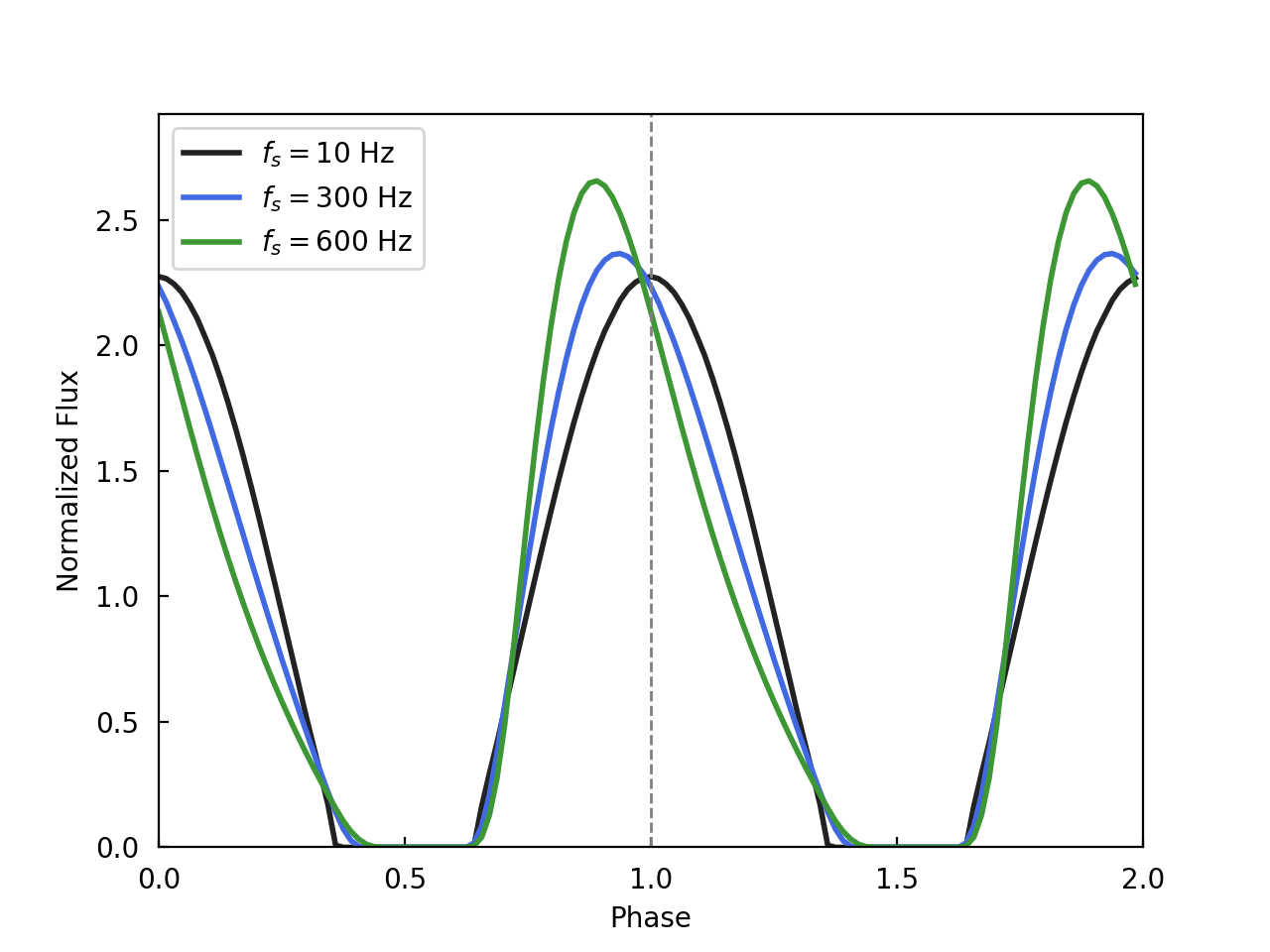}
\caption{The bolometric pulse profile generated for a neutron star with a dipole magnetic field but only one emitting polar cap, for various spin frequencies. The magnetic inclination is $\zeta=75\degree$, and the observer inclination is $\theta_o=90\degree$. As the spin frequency increases, the combination of Doppler and time-delay effects introduces substantial skewness to the pulse profile.}
\label{fig:fig2}
\end{figure}

In the interest of simplicity and clarity of comparison to previous dipole models, we calculate lightcurves from neutron stars with dipolar magnetic fields by considering only isotropic surface emission, i.e., setting the beaming factor $h=0$ so that $\mathcal{B}=1$. In order to understand our results below, we first consider the lightcurve due to a single polar cap; the full lightcurve is then just the sum of the contribution from each polar cap separately. Figure~\ref{fig:fig2} shows the lightcurve from a single polar cap on a neutron star with magnetic inclination $\zeta=75\degree$, spinning at three different frequencies, and observed at an inclination angle of $\theta_o=90\degree$. As has been shown earlier, the combination of Doppler and time-delay effects introduces a skewness to the lightcurve, making the rising (left) edge of the curve steeper and the waning (right) edge shallower~\citep{Braje2000}.

\begin{figure*}[t]
\centering
\subfigure{\label{fig:a}\includegraphics[width=0.32\textwidth]{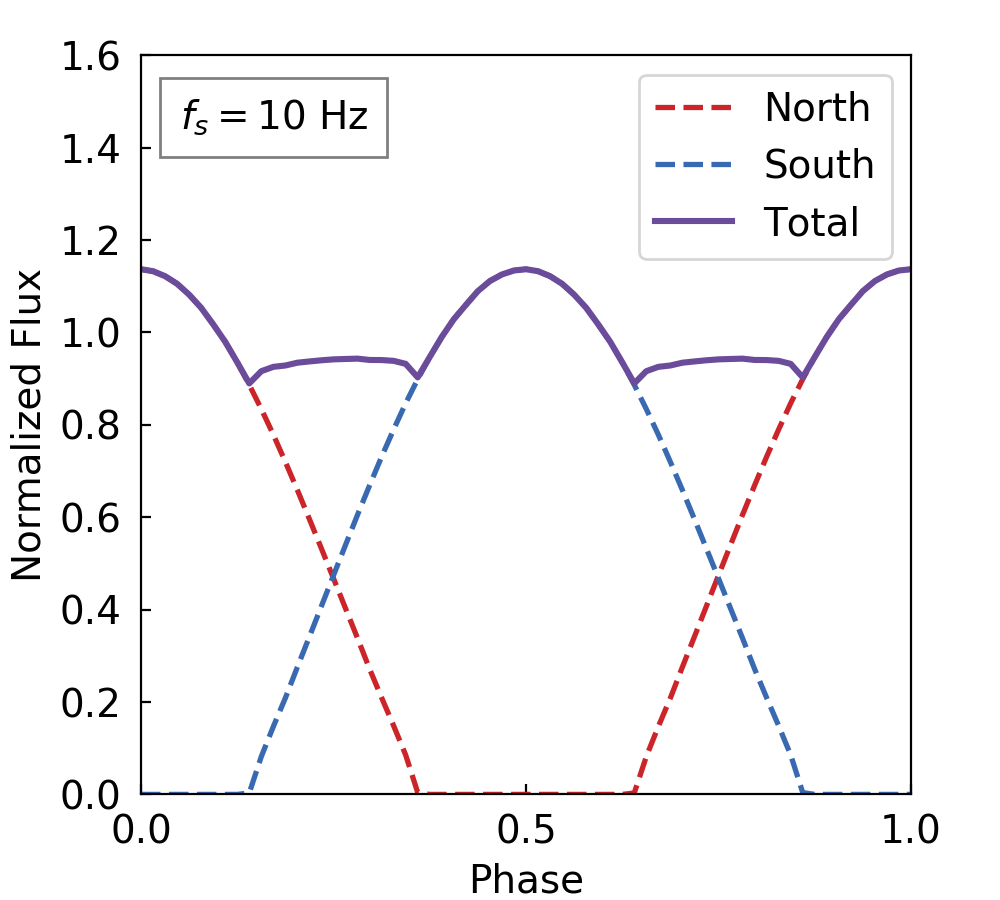}}
\subfigure{\label{fig:b}\includegraphics[width=0.32\textwidth]{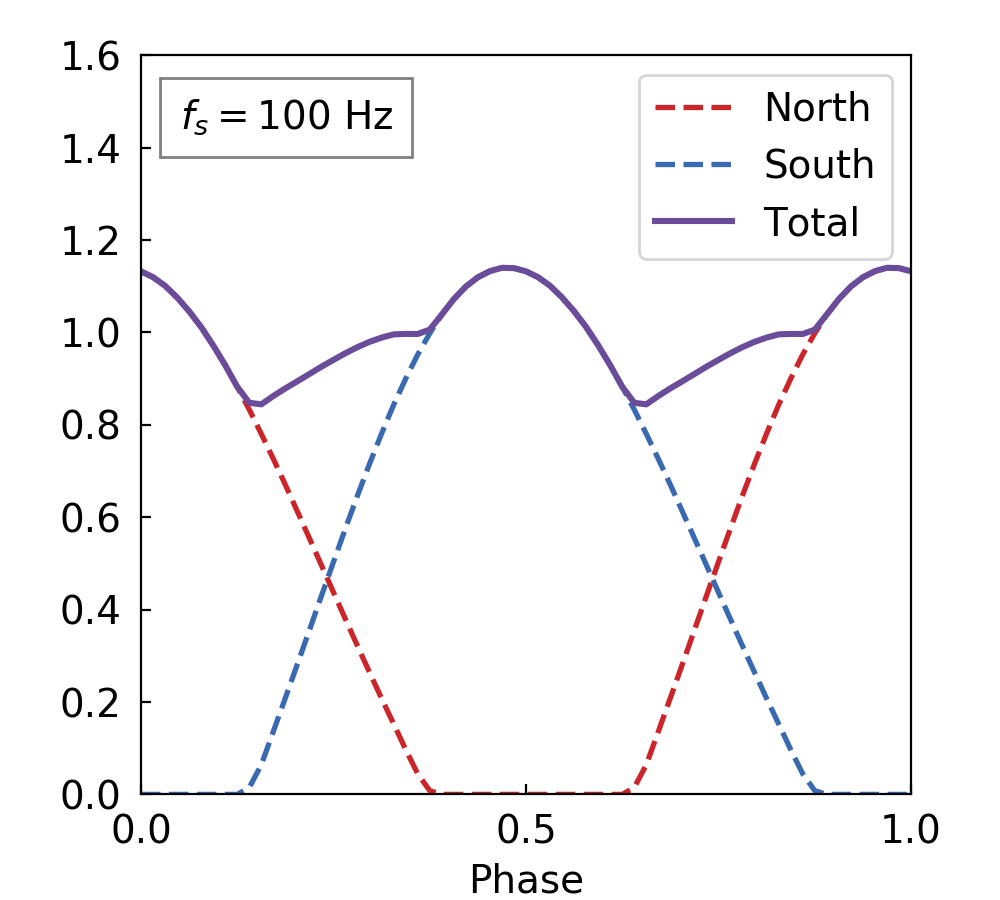}}
\subfigure{\label{fig:c}\includegraphics[width=0.32\textwidth]{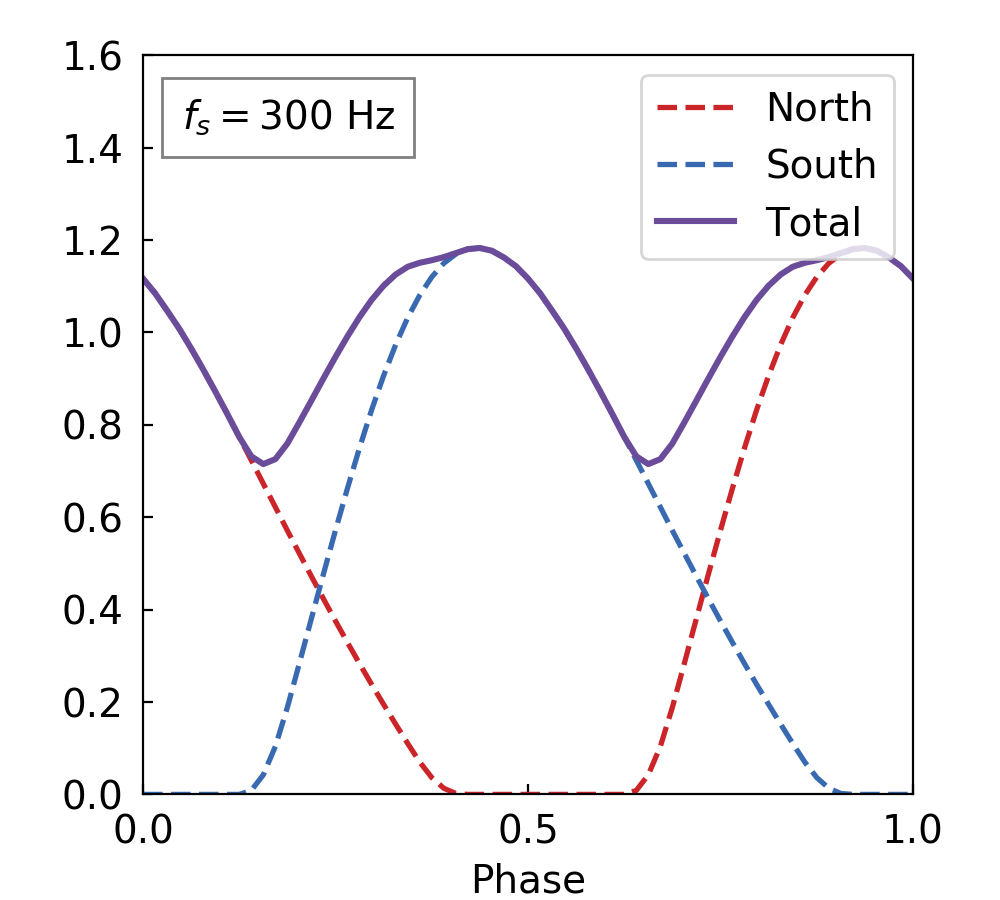}}
\setlength{\belowcaptionskip}{10pt} 
\caption{Same as Figure~\ref{fig:fig2}, but for emission arising from two polar caps. The increasing skewness of the lightcurves arising from each polar cap leads to asymmetric troughs between the pulses. 
\label{fig:fig3}}
\end{figure*}

When two of these skewed lightcurves are added together, $180\degree$ out of phase, to generate the lightcurve of a neutron star with two polar caps in the pure dipole field configuration, they give rise to the asymmetric troughs shown in Figure~\ref{fig:fig3}. The most striking features are the increased complexity in the lightcurves arising from such simple geometric polar caps and the fact that their time-reversal symmetry is broken.

In order to explore the dependence of the simulated lightcurves on the shape and temperature profile across the polar caps, we compare the results from three different prescriptions: 

\begin{enumerate}[label=(\Alph*)]
    \item constant temperature over the circular polar cap (this is the geometry that has been employed in most previous analyses);
    \item constant temperature over just the spacelike-current region; 
    \item the phenomenological model described in \S\ref{sec:method}, with a temperature $T \propto |\vec{J}|^{1/4}$ within the spacelike-current region (Eq.~\ref{eq:TJ4}).
\end{enumerate}

\begin{figure}
\centering  
\subfigure{\includegraphics[width=0.94\linewidth]{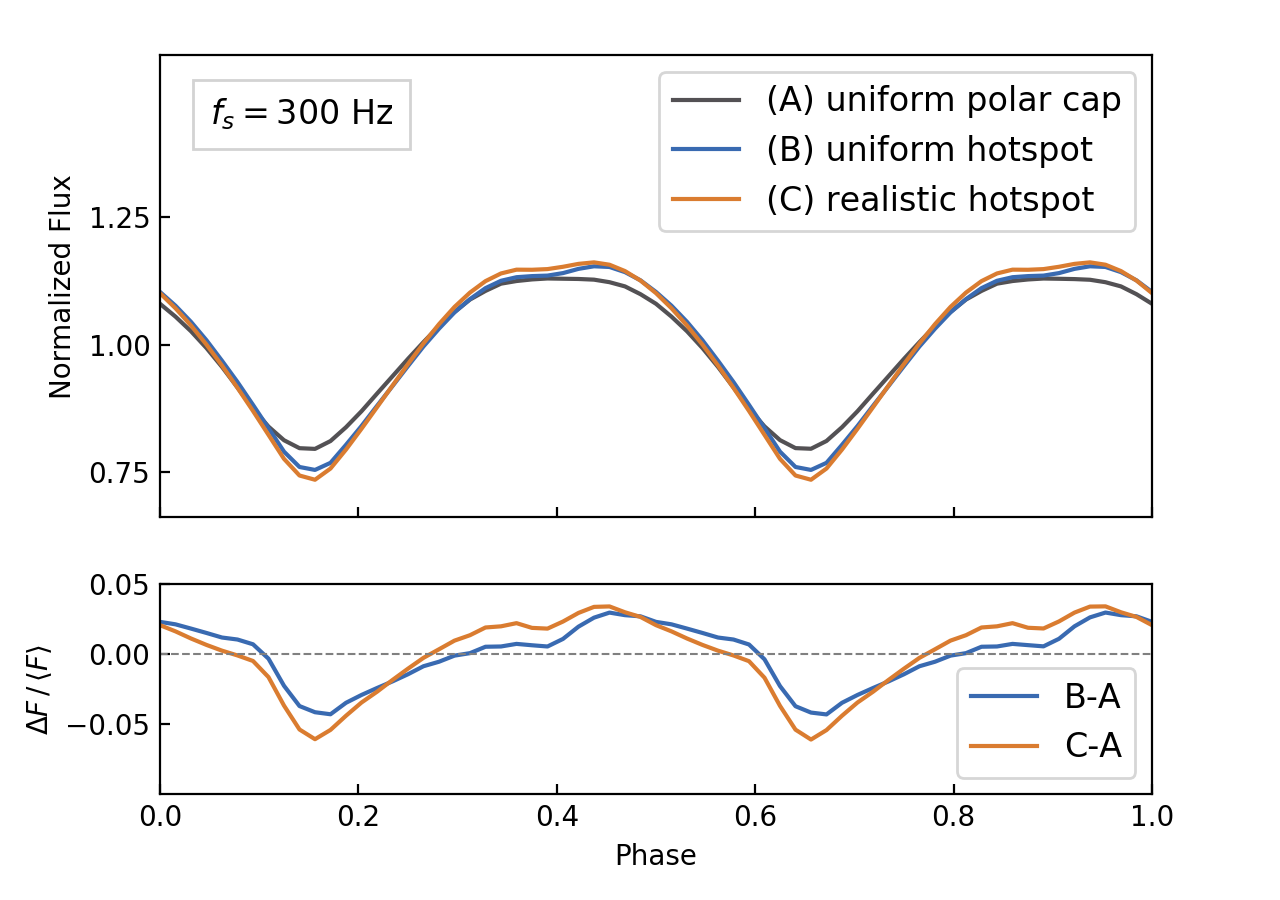}}
\subfigure{\includegraphics[width=0.94\linewidth]{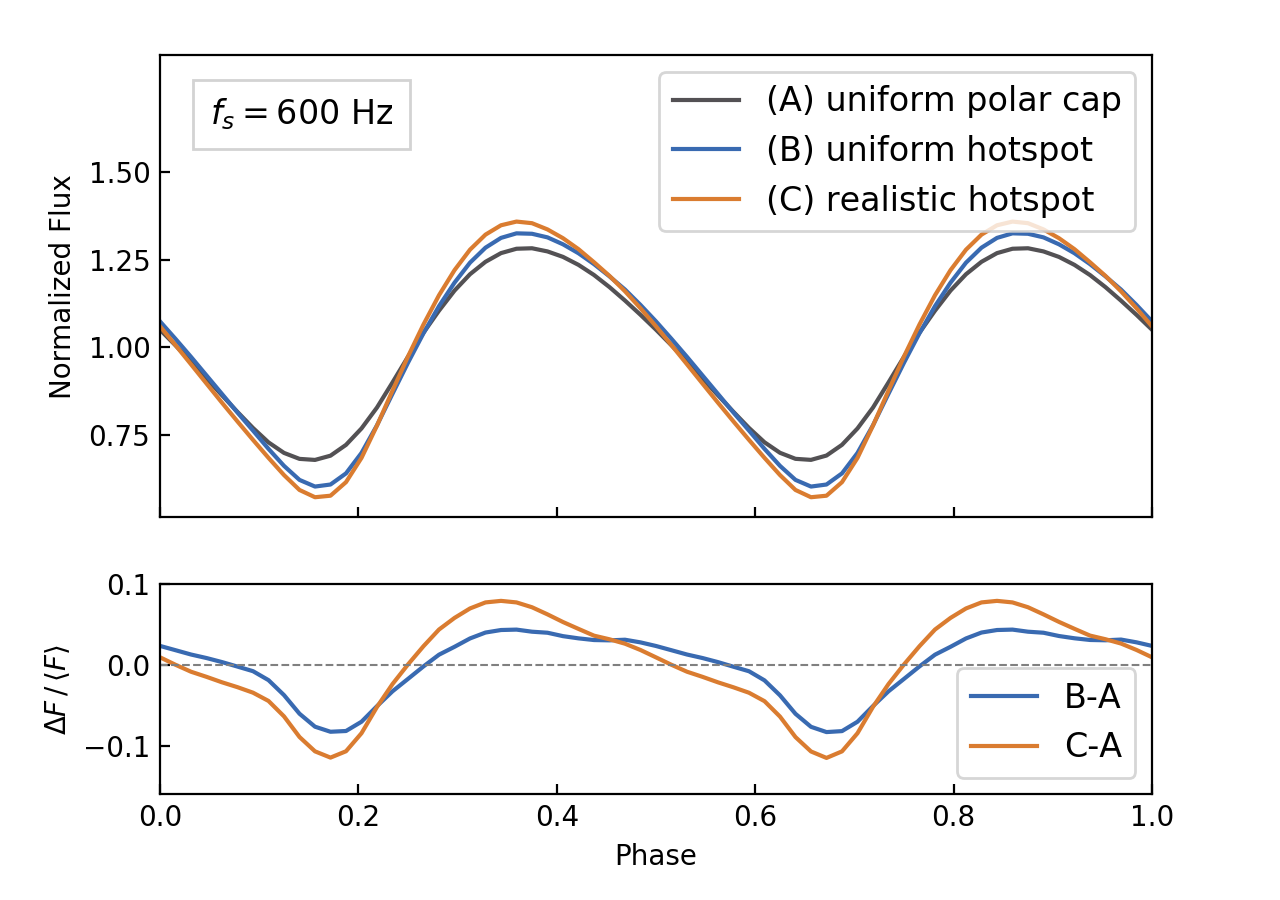}}
\caption{Pulse profiles from neutron stars with dipolar magnetic fields, for three different prescriptions of the shape and temperature profile of the emitting regions (see \S\ref{sec:dipole} for details).  In all cases, the magnetic inclination is $\zeta=60\degree$, the observer inclination is $\theta_o=90\degree$, and the spin frequency is set to {\em (top)\/} 300~Hz and {\em (bottom)\/} 600~Hz. The bottom panels of both figures show the fractional difference between the pulse profiles calculated for the different prescriptions. The detailed properties of the polar caps affect the resulting pulse profiles at the $\sim 5-10$\% level and may mask rotational effects that are of the same order. 
\label{fig:fig4}}
\end{figure}

\begin{figure}
\vspace{-10.8pt}
\centering  
\subfigure{\includegraphics[width=0.94\linewidth]{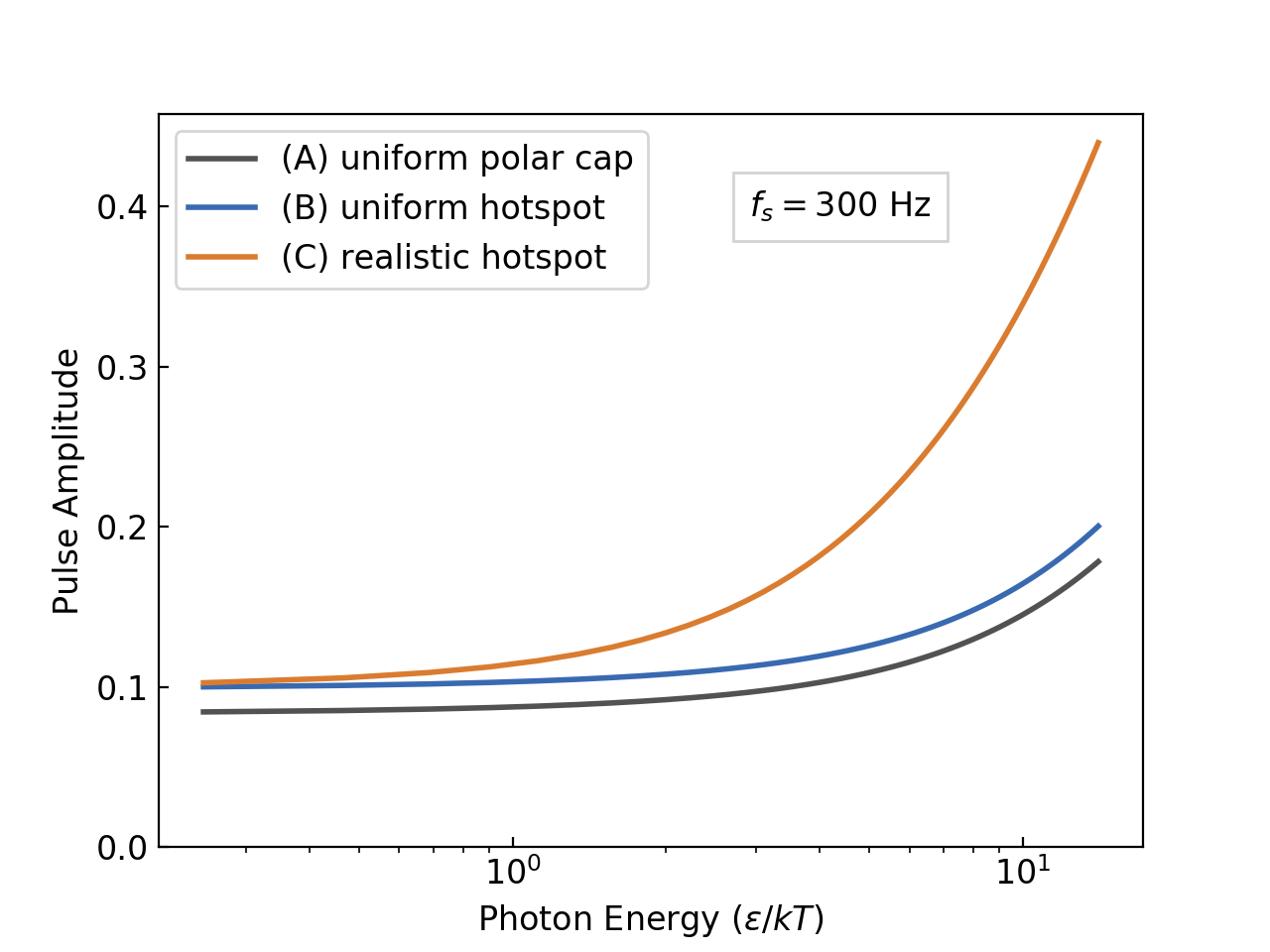}}
\subfigure{\includegraphics[width=0.94\linewidth]{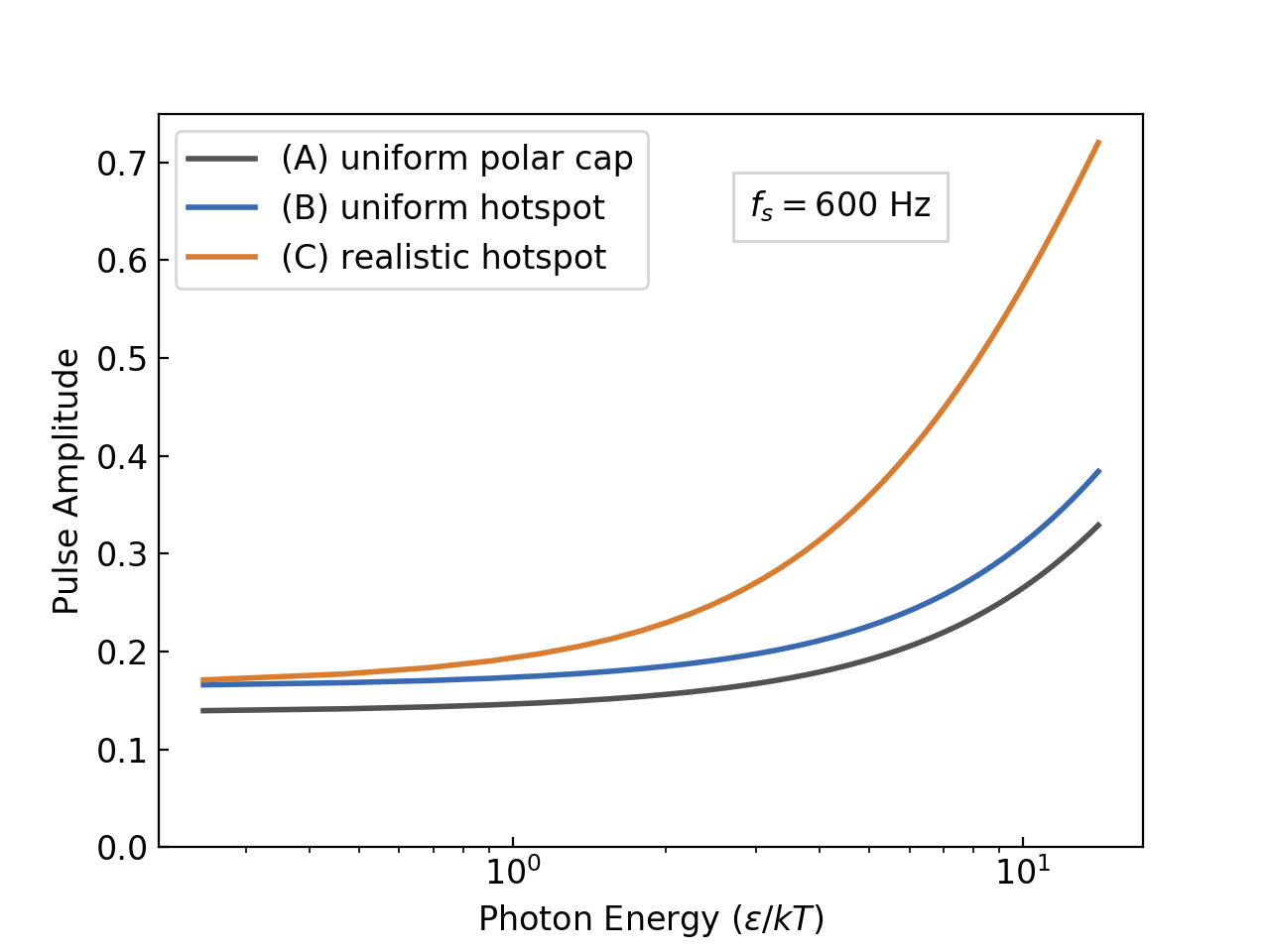}}
\caption{Pulse amplitude as a function of photon energy, for the configurations shown in Figure~\ref{fig:fig4}. For the constant-temperature polar caps, the weak dependence is the result of Doppler effects. The larger amplitudes and stronger dependence on photon energy for the realistic polar caps arises from the presence of temperature variation on the polar caps.}
\vspace{10pt}  
\label{fig:fig5}
\end{figure}

\begin{figure*}[t] 
\centering
\includegraphics[width=\textwidth]{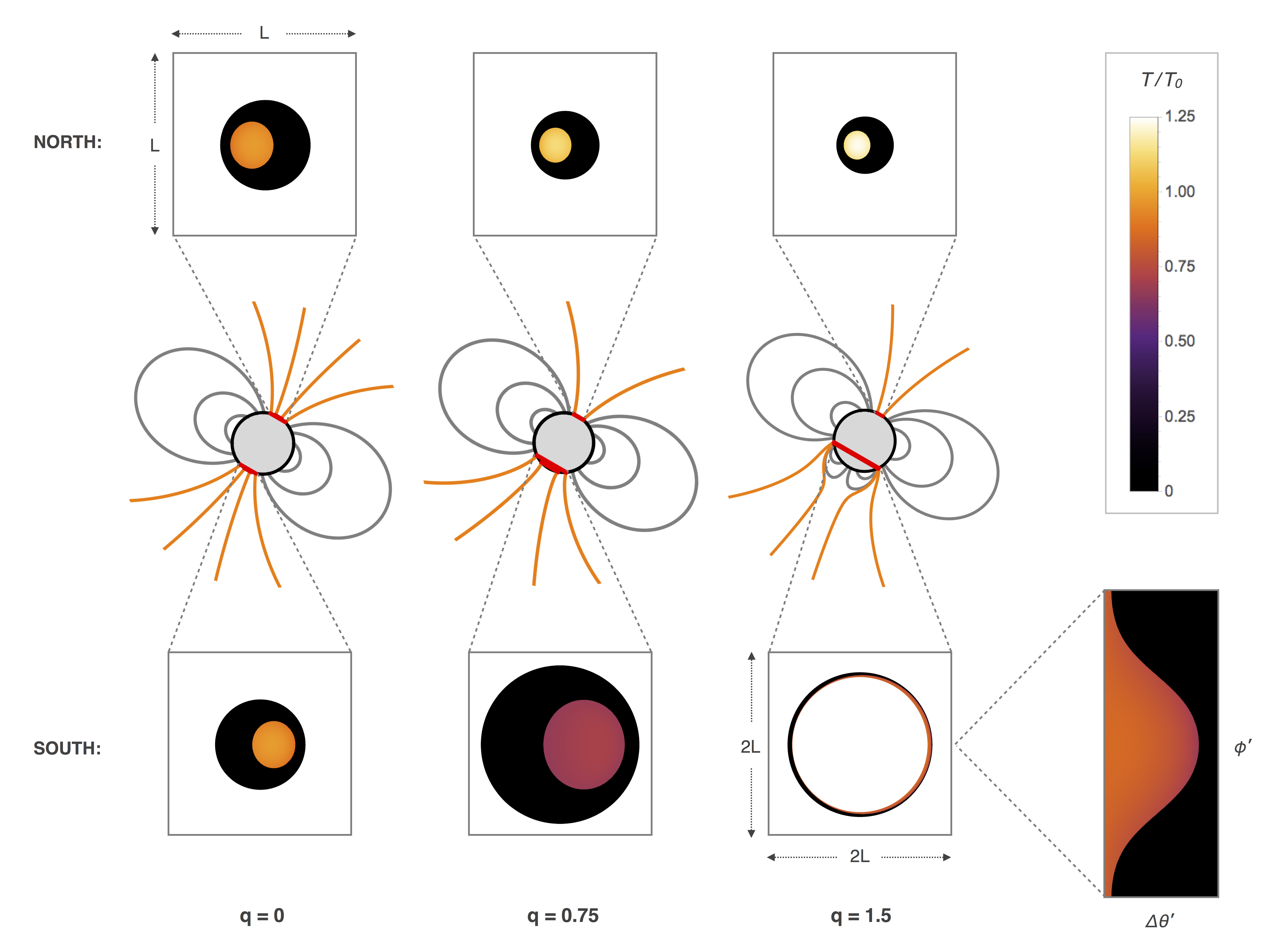}
\caption{Composite diagram depicting magnetic field configurations and temperature maps of the corresponding polar caps for a neutron star with a magnetic inclination $\zeta=30\degree$. The middle row shows the magnetic field configuration, with current flow along open field lines (orange) hitting the polar caps (red). The spin axis is oriented vertically. The top and bottom rows show the north and south polar caps respectively. As the relative magnitude of the quadrupole moment evolves from $q=0$ (left) to $q=1.5$ (right), we see that the northern cap shrinks and grows hotter, while the southern cap expands and cools. For $q=1.5$, the southern polar cap is a ring. A second insert (bottom right) shows the detailed structure of the ring as a plot in $(\theta',\phi')$, with $\phi'$ running vertically from 0 to $2\pi$. The polar caps are shown via parallel projection (i.e., how they would look in the Newtonian case with no lensing). For our fiducial parameters, the boxes bounding the polar caps all have a width $L$ equal to half the diameter of the star, except for the final box of width $2L$ showing the ring. For these parameters the ring has as angular size of $\Delta \theta' \simeq 5\degree$.
\label{fig:fig6}}
\vspace{15pt}
\end{figure*}

Figure \ref{fig:fig4} compares the normalized lightcurves for two neutron stars, one with moderate spin (300 Hz) and one with high spin (600 Hz), calculated with these three different phenomenological prescriptions. The detailed shape and temperature profiles of the polar caps 
introduce complexity to the predicted lightcurves at amplitudes that increase with the neutron-star spin frequency and can
reach levels as large as $\sim 5-10$\%. A large part of this complexity is caused by the range of temperatures that are present on the more realistic polar caps. As the star rotates, the relative projection of different parts of the polar cap that have different temperatures evolves with time, causing the observed spectrum to also change with time. The net result is a pulse amplitude from realistic polar caps that is larger than that of a polar cap with constant temperature~\citep[cf.\/][]{Baubock2015}. 

This spectral evolution with pulse phase also results in a pulse amplitude that depends on photon energy.  Figure~\ref{fig:fig5} shows the RMS fractional pulse amplitude\footnote{The RMS fractional amplitude is defined as the square root of the time-average of the squared deviation from the mean flux.} as a function of photon energy for the configurations shown in Figure~\ref{fig:fig4}. The weak dependence on photon energy seen in the pulse amplitudes of configurations with constant temperatures across the polar caps is caused entirely by rotational Doppler effects~\citep{Psaltis2014b}. The larger amplitudes and stronger dependence on photon energy seen in the simulations with realistic polar caps is a direct outcome of the presence of regions with different temperatures across each polar cap.

The increased amplitudes and strong photon energy dependence of the pulse profiles generated with realistic polar caps competes with and may mask the consequences of the rotational Doppler effects, which are qualitatively similar. If we were to attempt to fit the pulse profiles shown in Figures~\ref{fig:fig4} and \ref{fig:fig5} using models with circular, constant temperature polar caps, we would be forced to increase artificially the rotational effects in order to account for the large amplitudes. For a neutron star of 
known spin frequency and observer orientation, the only way to achieve the latter is to increase artificially the stellar radius.
As a result, modeling pulse profiles with circular, constant temperature emitting regions introduces the risk of biasing the inferences of the neutron-star radii.

\section{Pulsars with Quadrudipolar Magnetic Fields} \label{sec:quadru-dipole}

We now consider the effect on the simulated pulse profiles of adding a quadrupole moment to the magnetic field. This amounts to including the next term in the multipole expansion of the Euler potential,
\begin{align}
\begin{split}
     \alpha(r) &= B_1 R^2 \left( \frac{R^{>}_1(r)}{R^{>}_1 (R)} \right) \sin^2{\theta'} \, + \\
     &\quad B_2 R^2 \left( \frac{R^{>}_2(r)}{R^{>}_2(R)} \right) \cos{\theta'}\sin^2{\theta'},
\end{split}
\end{align}
where the radial functions are given by 
\begin{subequations} \label{eq:R2}
    \begin{align}
        R^{>}_1(r) &= \left(\frac{3}{2r}\right)\frac{3-4f+f^2+2\log{f}}{(1-f)^3}, \\
        R^{>}_2(r) &= \left(\frac{10}{3r^2}\right)\frac{17-9f-9f^2+f^3+6(1+3f)\log{f}}{(1-f)^5}.  
    \end{align}
\end{subequations}
We refer to this family of dipole-plus-quadrupole fields as ``quadrudipoles'' and parameterize them by the ratio $q \equiv B_2/B_1$, with $q=0$ corresponding to the pure dipole.

The addition of a magnetic quadrupole moment changes both the current density and the size and geometry of the polar caps, as shown in Figure~\ref{fig:fig6}. For $q<1$, the effect of the magnetic quadrupole moment is to spread out the field lines in the southern hemisphere and contract those in the north.\footnote{We take north to point in the direction of the spin vector according to the right-hand rule.} As a result, the northern cap shrinks and grows hotter with increasing quadrupole, while the southern cap expands and cools. When $q > 1$, the southern cap is no longer circular but becomes an annular strip encircling the star. When the magnetic field is substantially inclined from the spin axis, a large temperature gradient emerges along this ring (see bottom right insert in Figure~\ref{fig:fig6}), since only part of this region has spacelike four-current.

\begin{figure}
\centering
\includegraphics[width=\linewidth]{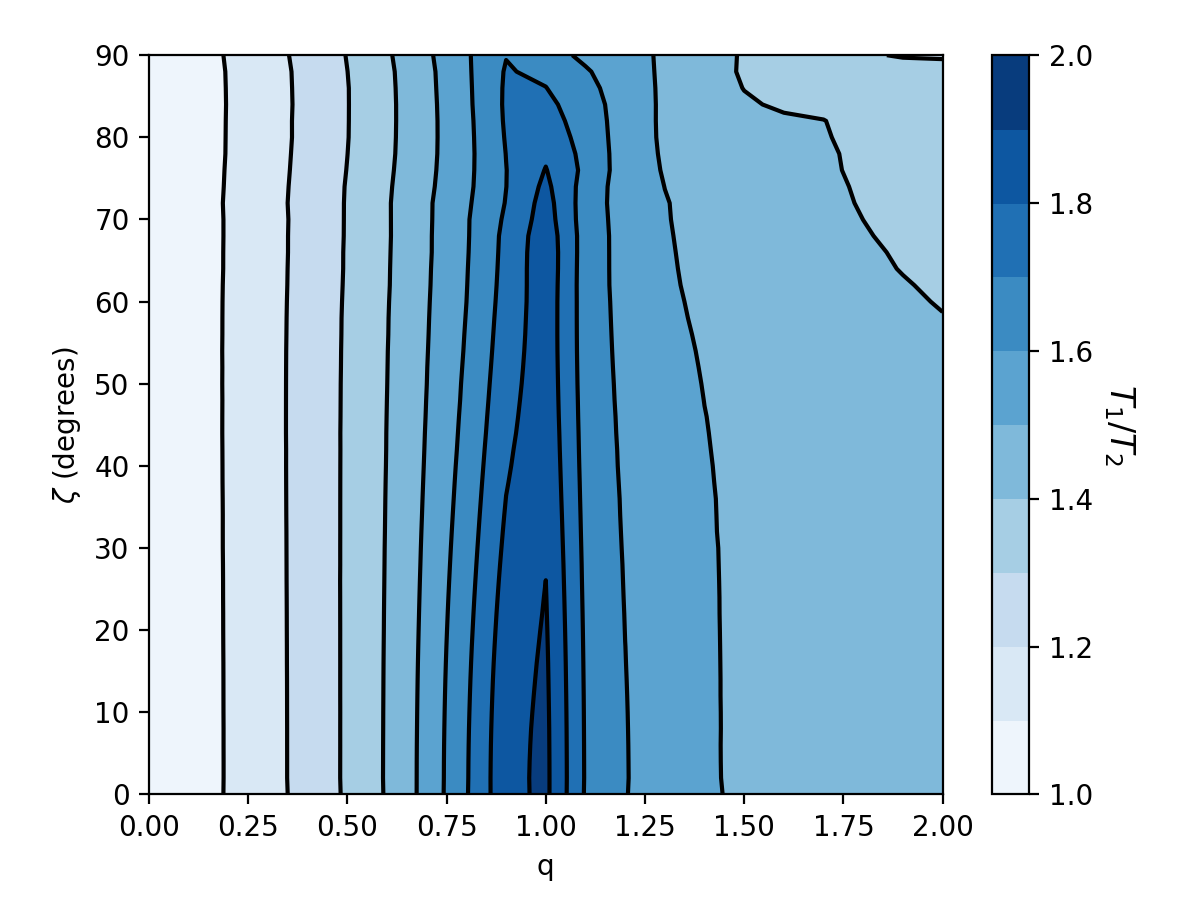}
\caption{Contours of constant ratio between the average temperatures of the northern and southern emitting regions of pulsars with different magnetic inclinations ($\zeta$) and quadrupole strengths ($q$). The temperature ratio reaches a maximum at $q=1$, the critical point at which the shape of the southern emitting region evolves from a spot to a ring. \label{fig:fig7}}  
\end{figure}

An important consequence of the quadrudipole model is that the northern and southern emitting regions have different effective temperatures. We can quantify this difference by calculating the ratio of the average temperatures of the two emitting regions as a function of the magnetic quadrupole moment $q$ and the magnetic inclination $\zeta$. Figure~\ref{fig:fig7} shows that this temperature ratio depends primarily on $q$, initially increasing with $q$ for $q \leq 1$. This is a direct result of the fact that the total current flow onto each polar cap is the same. As the field lines spread out with increasing $q$, the surface area of the southern cap grows and the current density (which determines the temperature) decreases. 
A quadrudipole field with $q=1$ marks the critical point at which the southern emitting region transitions from being a spot to a ring. Beyond this point, the area of the southern region decreases with increasing $q$, causing its temperature to rise. 

The quadrudipole configuration clearly breaks antipodal symmetry because the emitting regions in general have dramatically different shapes. The one symmetry that is maintained is reflection symmetry across the $\Omega \! - \!\mu$ plane, i.e. the plane determined by the spin ($\vec{\Omega}$) and magnetic dipole ($\vec{\mu}$) vectors. This reflection symmetry guarantees that the phases of greatest flux \textit{emitted} from each hotspot are still $180\degree$ apart. However, when the full ray-tracing calculation is carried out --- including Doppler shift, time-delays, and lensing --- the phases of peak flux \textit{observed} are no longer evenly spaced. The quadrudipole configuration makes possible additional features in the lightcurve that are not seen with the antipodal configuration of the pure dipole. 

\begin{figure}
\centering  
\includegraphics[width=0.96\linewidth]{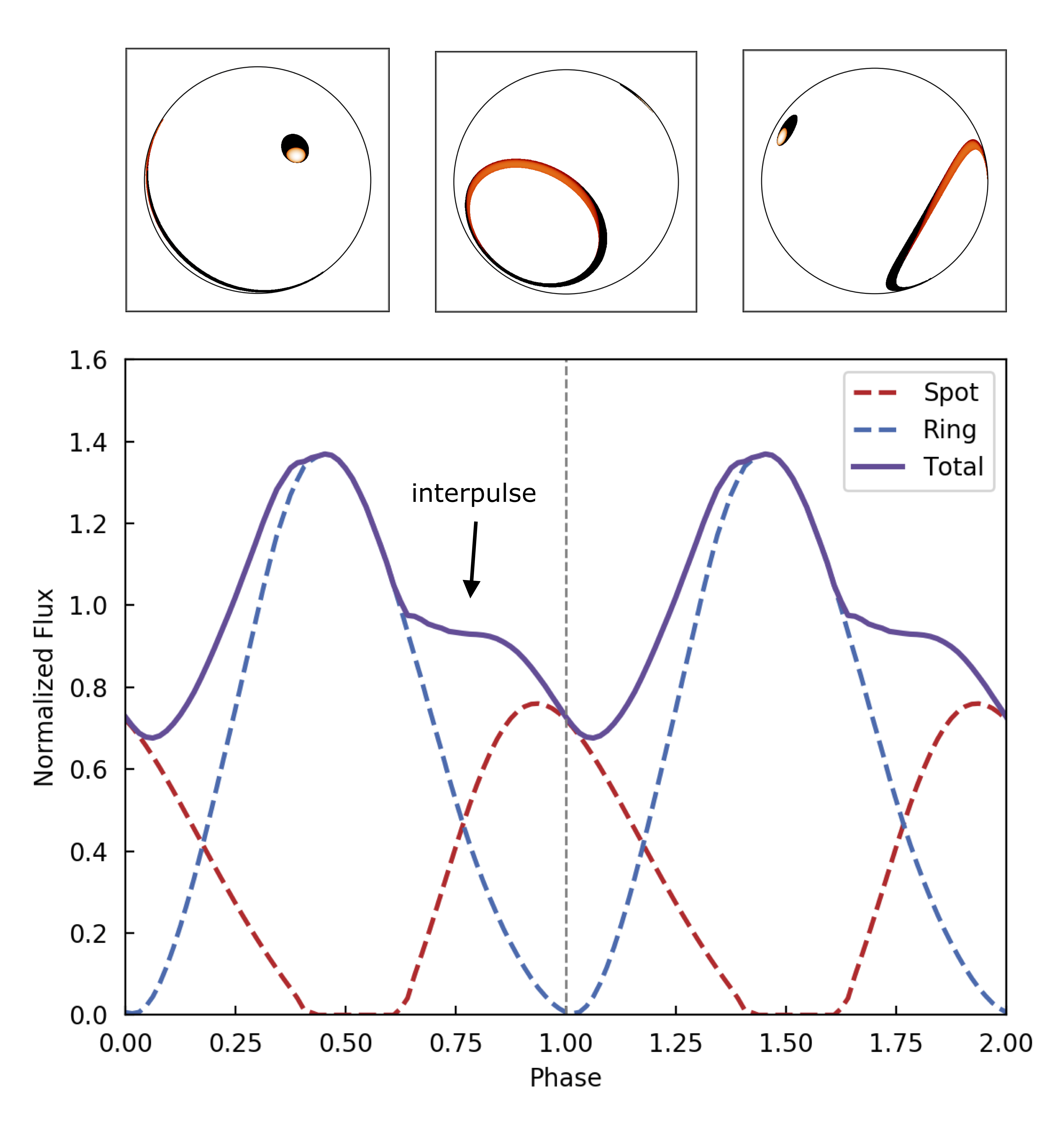}
\caption{{\em (Top)\/} Neutron-star snapshots at three equally-spaced phases in the rotation ($1/12$, $5/12$, and $9/12$ of a period) and {\em (Bottom)\/} pulse profile for a star with a quadrudipole magnetic field and $q=1.5$. The magnetic inclination is $\zeta=60\degree$ and the observer inclination is $\theta_o=80\degree$. The different geometries of the northern and the southern emitting regions generate an offset interpulse even for a purely axisymmetric magnetic field.
\label{fig:fig8}}
\vspace{-2mm}
\end{figure}

Figure~\ref{fig:fig8} shows the simulated lightcurve from a pulsar with a quadrudipole magnetic field ($q=1.5$) for a particular choice of parameters. The interesting feature of this pulse profile is the appearance of an `interpulse', which is offset in phase from the main peak by less than $180\degree$. This interpulse, which we might have naively identified with an off-axis hotspot, in fact arises because the emission regions from the northern spot and the southern ring are not antipodally symmetric, and their contributions to the pulse profile are affected differently when subject to the Doppler shift and time delays. 
Overall, we find temperature ratios as large as $\sim 2.0$ between the two emitting regions (Figure~\ref{fig:fig7}) and pulse profiles with offset interpulses (Figure~\ref{fig:fig8}), potentially providing a natural explanation to the multi-temperature blackbody models and non-antipodal emission geometries that are required to fit the observed spectra of millisecond pulsars~\citep[see, e.g.,][]{Bogdanov2013}.

\section{Application to X-ray Observation \\ of PSR J0437$-$4715} \label{sec:J0437}

PSR~J0437$-$4715 is the nearest known rotation-powered millisecond pulsar and the best candidate for measuring neutron-star properties via pulse profile modeling \citep{Arzoumanian2014}. It has a spin frequency $f_{\rm s} = 174 \rm{Hz}$ and orbits a white dwarf companion with an orbital period of 5.7 days. Pulsar timing measurements yield a pulsar mass of $M=1.44 \pm 0.07 M_\odot$ \citep{Reardon2016}. 

The X-ray properties of this pulsar have posed significant challenges to modeling \citep{Bogdanov2013}. As discussed earlier, fitting the X-ray spectra of the surface emission requires multi-temperature blackbody models. The lightcurve itself exhibits an interpulse that is not evenly-spaced between the main pulses, but is rather shifted forward by roughly 20 degrees. In order to model this lightcurve, \cite{Bogdanov2013} employed an ``offset dipole'' model, in which the magnetic field is dipolar but the dipole is not located at the geometric center of the star. This gives rise to emitting regions that are not antipodal. Nevertheless, the shapes of the hotspots are assumed to be circular. Employing this model to fit the observations requires four parameters in addition to the ones required for the model we explore here: angular offsets $\Delta\theta$ and $\Delta\phi$ that specify the location of one of the hotspots with respect to the other, the ratio of the two hotspot radii, and, independently, the ratio of their temperatures.

\begin{figure}
\centering
\includegraphics[width=0.96\linewidth]{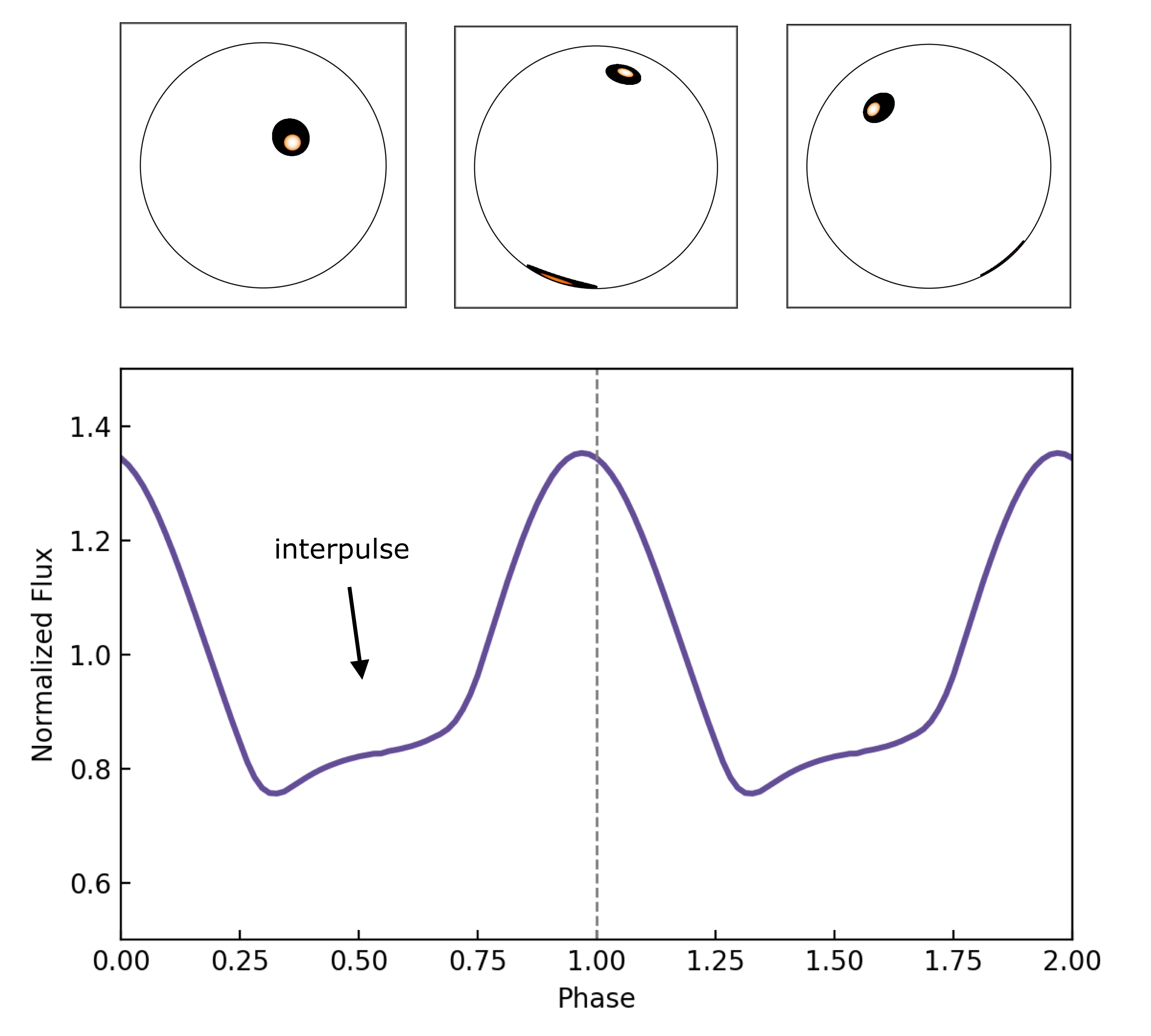}
\caption{{\em (Top)\/} Neutron-star snapshots (same phases as Figure \ref{fig:fig8}) and {\em (Bottom)\/} pulse profile for a pulsar with the following parameters: $M=1.4M_\odot$, $R=12\textrm{km}$, $f_s=174\textrm{Hz}$, $\zeta = 30\degree$, $\theta_o = 42\degree$, $q=0.5$, and a beaming factor of $h=0.4$. The interpulse on the leading edge of the main pulse is suggestive of the observed lightcurve of PSR~J0437$-$4715, which has previously been fit with a more complicated offset dipole model \citep{Bogdanov2013}.
\label{fig:fig9}}
\end{figure}

The quadrudipole model offers the potential of reproducing the key qualitative features of PSR~J0437$-$4715 in a more satisfactory way by introducing the single parameter $q$ (the ratio of the aligned magnetic quadrupole to dipole moments) while retaining axisymmetry. Figure~\ref{fig:fig9} shows a model lightcurve that is suggestive of a potential fit to PSR~J0437$-$4715. The underlying magnetic field structure is quadrudipolar with $q=0.5$. \\ For the purposes of this plot, we have assumed a magnetic inclination of $\zeta=30\degree$ and an observer inclination of $\theta_o=42\degree$, which is the value inferred for this pulsar from the inclination of its binary orbit~\citep{Reardon2016}.  In order to generate a large pulse amplitude, we have also used non-isotropic beaming, as expected for the atmospheres of pulsars that are bombarded by magnetospheric charges~\citep{Baubock2019}, and chose $h=0.4$. In the simulated pulse profile, the main peak corresponds to the northern emitting region, which is seen almost head-on at its peak intensity. The interpulse is due to the southern emitting region, seen at a glancing angle. 

This quadrudipole model offers several advantages over the offset dipole. First, we only need to add a single new parameter to the classic dipole model, without breaking axisymmetry in the magnetic field. Second, a natural consequence of the quadrudipole model is that the hotspot with the smaller surface area will also have the higher temperature. This is consistent with the relationship found in the multi-temperature blackbody fits of the data from PSR~J0437$-$4715. Finally, the appearance of the offset interpulse is robust enough that we are also able to set the observer angle to $\theta_o = 42\degree$, as is inferred for this source.

This quadrudipole model offers several advantages over the offset dipole. First, we only need to add a single new parameter to the classic dipole model, without breaking axisymmetry in the magnetic field. Second, a natural consequence of the quadrudipole model is that the hotspot with the smaller surface area will also have the higher temperature. This is consistent with the relationship found in the multi-temperature blackbody fits of the data from PSR~J0437$-$4715. Finally, the appearance of the offset interpulse is robust enough that we are also able to set the observer angle to $\theta_o = 42\degree$, as is inferred for this source.

Performing a detailed fit of our model to the data would require taking into account the effects of interstellar absorption, the response of the detector, as well as the pulsed magnetospheric emission that contaminates the measurement of the surface emission~\citep{Guillot2016}.  We expect to report on these issues in the future.

\section{Conclusions}

We conclude by summarizing our key results and discussing some limitations of the present work as well as promising future directions.  
We have shown that the shape and temperature profile across the pulsar hotspots plays an important role in lightcurve modeling that cannot be ignored, and that significant complexity in lightcurve features is possible simply with the addition of a quadrudipolar magnetic field.

This work is based on an analytic model of the pulsar magnetosphere that is guided by state-of-the-art numerical simulations. We can expect refinements of the magnetospheric structure and currents as numerical simulations improve. For example, volume return currents, which we ignore in our present model, may also play a role in atmospheric heating \citep{Timokhin2013,Philippov2018}. However, we expect that the basic qualitative results presented here will remain unchanged and that this approach will aid in measuring neutron star parameters and ultimately help tighten constraints on the neutron-star equation of state.

\section*{Acknowledgments} 

We thank Sasha Philippov and Deepto Chakrabarty for helpful discussions and valuable comments on the manuscript. We gratefully acknowledge support from NASA grant NNX16AC56G and NSF grant PHY-1752809.

\bibliography{hotspots}

\end{document}